\documentclass[12pt,preprint]{emulateapj}
\slugcomment{Submitted to ApJ Letters (Revised Version including Referee's Comments)}

\shorttitle{Silicate Feature in A DLA}
\shortauthors{Kulkarni et al.}

\begin{document}


\title{9.7 $\mu$ m Silicate Absorption in A Damped Lyman-$\alpha$ Absorber at $z=0.52$}

% As in the title, use \\ to force line breaks.
\author{Varsha P. Kulkarni\altaffilmark{1}, Donald G. York\altaffilmark{2,3}, 
Giovanni Vladilo\altaffilmark{4}, Daniel E. Welty\altaffilmark{2}}

\altaffiltext{1}{Department of Physics and Astronomy, University of South Carolina, 
Columbia, SC 29208; E-mail: kulkarni@sc.edu}
\altaffiltext{2}{Department of Astronomy and Astrophysics, University of Chicago, 
Chicago, IL 60637}
\altaffiltext{3}{Also, Enrico Fermi Institute}
\altaffiltext{4} {INAF, Osservatorio Astronomico di Trieste, Trieste, Italy}



\begin{abstract}
We report a detection of the 9.7 $\mu$m silicate absorption 
feature in a damped Lyman-$\alpha$ (DLA) system   
at $z_{abs} = 0.524$ toward AO0235+164, using the Infrared Spectrograph (IRS) onboard 
the {\it Spitzer Space Telescope}. The feature shows a broad shallow profile 
over $\approx$ 8-12 $\mu$m in 
the absorber rest frame and appears to be 
$> 15 \, \sigma$ significant in equivalent width. The feature is fit 
reasonably well by  
the silicate absorption profiles for laboratory amorphous olivine 
or diffuse Galactic 
interstellar clouds. To our knowledge, 
this is the first indication of 9.7 $\mu$m silicate absorption in a 
DLA.  We discuss potential implications of this 
finding for the nature of the dust in quasar absorbers. 
Although the feature 
is relatively shallow ($\tau_{9.7} \approx 0.08-0.09$), it is 
 $\approx 2$ times deeper than expected from extrapolation of the $\tau_{9.7}$ {\it vs.} 
$E(B-V)$ relation known for diffuse Galactic interstellar clouds. Further studies 
of the 9.7 $\mu$m silicate feature in quasar absorbers will open a 
new window on the dust in distant galaxies. 

\end{abstract}


\keywords{Quasars: absorption lines--ISM:dust}



\section{Introduction}

Damped Lyman-alpha (DLA) absorption systems in quasar spectra dominate the 
neutral gas content in galaxies and offer venues for studying the evolution 
of metals and dust in galaxies. Recent observations, however, suggest that the 
majority of DLAs have low metallcities at all 
redshifts studied ($0 \lesssim z \lesssim 4$), with the mean 
metallicity reaching at most $\approx 10-20 \%$ 
solar at the lowest redshifts (see, e.g., Prochaska et al. 2003; Kulkarni et al. 
2005, 2007; P\'eroux et al. 2006; and 
references therein). These results appear to contradict the predictions 
of a near-solar global mean interstellar metallicity of galaxies at $z \sim 0$ in most chemical evolution 
models based on the cosmic star formation history inferred from galaxy imaging surveys 
such as the Hubble Deep Field (HDF) (e.g., Madau et al. 1996). 
Furthermore, for a large fraction of the DLAs, the 
SFRs inferred from emission-line imaging searches fall far below the global 
predictions (e.g., Kulkarni et al. 2006, and references therein). 

A possible explanation of these puzzles is that the current DLA samples are 
biased due to dust selection effects, i.e. that the more dusty and more metal-rich 
absorbers obscure the background quasars more, making them harder to observe 
(e.g., Fall \& Pei 1993; Boiss\'e et al. 1998; Vladilo \& P\'eroux 2005). DLAs 
are known to have some dust, 
based on both the (generally mild) depletions of refractory elements and the 
(typically slight) reddening of the background quasars (e.g., Pei et al. 1991; 
Pettini et al. 1997; Kulkarni et al. 1997). Combining $\sim 800$ 
quasar spectra from the Sloan Digital Sky Survey (SDSS), York et al. (2006b) 
found a small but significant 
amount of dust in absorbers at $1 < z < 2$,  with $E(B-V)$ 
of 0.02-0.09 for 9 of their 27 sub-samples (see also Khare et al. 2007). 
York et al. (2006b) also   
showed that the extinction in the composite spectra is best fitted by a Small Magellanic Cloud (SMC) curve (with
no 2175 {\AA} bump). Some recent 
studies suggest that dusty DLAs could hide as much as $17 \%$ of the total 
metal content at $z \sim 2$, and more at lower $z$ (Bouch\'e et al. 2005).  
To understand whether this is the case, and to understand the role of 
dust in quasar absorbers in general, it is essential to 
directly probe the basic properties of the dust. 

Recently, a small number of very dusty 
quasar absorbers have been discovered,  via various signatures 
of the dust in optical and UV observations: 
substantial reddening of the background quasars, large element depletions 
(e.g., for Cr, Fe), and/or a detectable 2175 {\AA} bump 
(e.g., Junkkarinen et al. 2004; Wang et al. 2004). 
It is not yet clear, however, whether the dust in these systems is similar to 
that in the Milky Way or SMC or LMC. 

 The 2175 {\AA} bump is generally, though not conclusively, attributed to 
carbonaceous grains. The silicate component of the dust, 
 believed to comprise $\approx 70 \%$ of the core mass of 
interstellar dust grains in 
the Milky Way (see, e.g., Draine 2003) has not yet been probed in quasar 
absorbers. A unique opportunity 
to study this important dust component is provided by the Spitzer IRS (Werner 
et al. 2004; Houck et al. 2004), which provides 
the spectral coverage, sensitivity, and resolution needed for the detection of the strongest 
of the silicate spectral features near 9.7 $\mu$m. The 9.7 $\mu$m feature, 
thought to arise in Si-O stretching vibrations, is 
seen in a wide range of Galactic and extragalactic environments 
(e.g., Whittet 1987 and references therein; Spoon et al. 2006; Imanishi 
et al. 2007). We have been carrying out 
an exploratory study of the silicate dust in quasar absorbers by 
searching for the 9.7 $\mu$m absorption feature with the Spitzer IRS. Here we report 
on the detection of the 
9.7 $\mu$m feature in one of the systems studied, while the remaining three systems observed recently will be reported 
in a separate paper (Kulkarni et al. 2007b, in preparation). 
 
\section{Observations and Data Analysis}
 The DLA at $z_{abs}=0.524$ (Junkkarinen et al. 2004) toward the blazar AO 0235+164  ($z_{em} = 0.94$) 
 offers an 
excellent venue for comparing dust in a distant galaxy with that in 
near-by galaxies. It has one of the largest H I column densities 
seen in DLAs (log $N_{\rm H I} = 21.70$) and shows 21-cm absorption 
(Roberts et al. 1976). It also shows X-ray absorption, consistent with a metallicity of 0.7 solar 
(Junkkarinen et al. 2004). Candidate absorber galaxies (much fainter 
than the blazar) within a few arcseconds 
from the blazar sightline have been detected (e.g., Smith et al. 1977; 
Yanny et al. 1989; Chun et al. 2006). This absorber is one of a very few 
DLAs producing appreciable reddening 
[$E(B-V)$ = 0.23 in the absorber rest frame] and detection of a strong 
broad 2175 \AA\ extinction bump (Junkkarinen et al. 2004).   Finally, this absorber is the 
only DLA with detections of 
several diffuse interstellar bands   
(Junkkarinen et al. 2004; York et al. 2006a). All of these data suggest that this 
absorber is very dusty and may contain molecular gas.

The observations were obtained with the Spitzer IRS on January 30, 2006  
(UT) as GO program 20757 (PI V. P. Kulkarni). IRS modules Short-Low 1 (SL1) and 
Long-Low 2 (LL2) 
were used to cover  7.5-21.4 $\mu$m in the observed frame (4.9-14.1 $\mu$m 
in  the DLA 
rest 
frame). The target was acquired with high-accuracy peakup 
using a near-by bright star.  The IRS standard staring mode was used, 
with 2-pixel slit widths of 3.6'' for SL1 and 
10.5'' for LL2. Integration times were 60 s $\times 8$ cycles for SL1 and 
120 s $\times 11$ cycles 
for LL2. For each cycle, observations were performed at both nod positions 
A and B (offset by 1/3 the slit length), 
so the total integration times were 960 s and 2640 s, respectively, for SL1 and LL2.   

The data were processed using the IRS S15.0 calibration pipeline 
(the latest version available at 
present), Image Reduction and Analysis Facility (IRAF\footnote{IRAF is distributed by the 
National Optical Astronomy Observatories, which are operated by the Association of 
Universities for Research in Astronomy, Inc. (AURA), under cooperative agreement with 
the National Science Foundation}), and Interactive Data Language (IDL). As detailed 
below,
the S15.0 pipeline yielded significant improvements for the reliable detection
and measurement of weak, broad features in our spectra. The pipeline performs 
a number of standard processing steps to produce the
basic calibrated data (BCD) files (see, e.g., the IRS Data Handbook 
at http://ssc.spitzer.caltech.edu/irs/dh).
Subtraction of the sky (mostly zodiacal light) 
was performed by subtracting the coadded 
frames at nod position B from those at nod position A, and vice versa. 
The 1-dimentional spectra were extracted from the 2-dimensional images using 
the Spitzer IRS Custom Extraction (SPICE) software using the default extraction 
windows, and flux calibrated using the standard S15.0 flux calibration files. 
The spectra 
from the two nod positions were averaged together, and the corresponding flux 
uncertainties calculated using both measurement uncertainties and ``sampling 
uncertainties'' between the two nod positions.

The absolute flux levels in the different IRS modules were 
scaled to match the continuum levels in the 
overlapping 
regions, using the bonus segment available in the LL2 images. There 
was no mismatch between the SL1 and LL2 flux levels; we used the SL1 data 
for $\lambda < 14.23 \,  \mu$m and LL2 data for $\lambda > 14.23 \, \mu$m. The data at $\lambda > 20 \, 
\mu$m bonus segment level had to be scaled up by $ 5.5 \%$ to match with the LL2 
data at $\lambda < 20 \, \mu$m. Fig. 1(a) shows the final merged spectrum of 
AO0235+164. The S/N achieved per unbinned pixel in the final spectrum, 
determined from rms fluctuations in the continuum regions, is $\approx 100$. 
The error bars denote 1 $\sigma$ uncertainties.  

The dashed line in Fig. 1(a) shows an estimate of the power-law continuum of the  
quasar. This line joins the observed continuum fluxes at 5.6 and 7.1 $\mu$m 
in the 
absorber rest frame and is extrapolated to the remaining wavelength region. These wavelengths 
are chosen to be in regions free of any other potential emission or absorption features 
(e.g., Imanishi et al. 2007). 
In principle, significant 9.7 $\mu$m emission at the quasar redshift
could affect continuum determination redward of the
suspected silicate absorption feature from the DLA.  However,
(a) our spectrum does not extend that far to the red, (b) the power
law provides a good fit to the continuum in our data, and (c) the
9.7 $\mu$m emission is not particularly strong in most quasars
(e.g., Hao et al. 2007).  

\section{Results}

The spectrum shown in Fig. 1(a) exhibits a broad
absorption feature between about 12.4 and 18.3 $\mu$m relative to the power law 
continuum.  The
flux decline from the continuum begins near the long wavelength end
of SL1 and continues smoothly into the LL2 data.  The broad feature is centered 
at 15.41 $\mu$m (10.11 $\mu$m in the DLA 
rest frame). The observed frame equivalent width is $0.31  \mu$m, with a  
1 $\sigma$ 
uncertainty of 0.014-0.020 $\mu$m, including  contributions from photon noise 
and continuum fitting uncertainties (Sembach \& Savage 1992).

We have performed several checks of our data analysis to see whether the observed broad feature 
could be an artifact. Since the possible silicate feature is  
broad and shallow, extending from the long wavelength end of SL1 through most of LL2, 
flux calibration and continuum fitting 
are critical issues. In the S14 pipeline version of these data, the possible 
silicate feature was somewhat 
stronger than in the S15 version. These differences are due to  
a low-level non-linearity problem in the S14 pipeline, which produces a $4 \%$ 
tilt in LL2 spectra and a $5 \%$ mismatch at the SL1/LL2 boundary. This problem 
has been eliminated in the  
S15 pipeline, and we find no mismatch at the 
SL1/LL2 boundary in the S15 data. 

The possible silicate feature does not show any visible signature of the
``teardrop'' feature known to exist near 14.1 $\mu$m in some SL1 data 
(see, e.g., 
the IRS data handbook). The 
beginning of decline in flux at the long-wavelength end of SL1 matches 
smoothly with the flux at the short-wavelength end of LL2 (which does not 
suffer from the teardrop problem). Our results do not change much 
even if the SL1 data are truncated at 14 $\mu$m to avoid the region potentially affected by the teardrop 
(the region 14-14.23 $\mu$m is a small 
fraction of the whole feature stretching out to 
18.3 $\mu$m in the observed frame). 

Inaccuracies in pointing (which can affect SL1 and LL2 
fluxes at the $\pm 1 \%$ level) also do not appear to be 
significant for our data. Based on an examination of the spectral 
images and the pointing difference keywords in the data file headers, 
the telescope pointing was accurate to within 0.09-0.11'' for LL2 and within 
0.22-0.29'' for SL1. 
Integrating a Gaussian intensity distribution from a point source 
with the Spitzer point spread function over the known slit dimensions 
($57 \arcsec \times 3.6 \arcsec$ for SL1, $168 \arcsec \times 10.5 \arcsec$ 
for LL2), we estimate that the effect of such an offset would be about 
$0.26 \%$ for SL1 and $0.05 \%$ for LL2, far too small to account for the 
observed feature. 

We also compared our results with IRS spectra from the literature for 
quasars without strong absorption systems (e.g., Sturm et al. 2006; Hao et al. 
2007), and did not find the broad absorption feature from our data in those 
other quasars. In fact, quasar spectra in general  show no silicate 
absorption, but rather (generally relatively weak) silicate emission at the 
quasar emission redshift. We also compared our IRS data for 
AO0235+164 with those for other targets in our study. The feature seen in 
AO0235+164 is not seen at the same observed wavelength in the other objects,  
suggesting that it is not an instrumental artifact. [In fact, 
in Kulkarni et al. 2007b (in prep.), 
we will report the possible detection of redshifted broad 9.7 $\mu$m 
silicate absorption in
other parts of the Spitzer spectral coverage toward other 
quasars.] 

Given the results of the above tests and the fact that the DLA toward 
AO0235+164 is already known to be dusty (from detection of 
2175 {\AA} bump and diffuse interstellar bands and reddening of the background 
quasar), it seems very likely that the feature detected is the broad 9.7 $\mu$m silicate feature arising in the absorber galaxy.

\section{Discussion}

The suggested silicate feature in the DLA toward AO0235+164 is 
relatively shallow/weak compared to 
the silicate features typically observed in Galactic interstellar material (ISM) 
because of the modest reddening and lower amounts of dust in quasar absorbers 
than in the Milky Way. Indeed, the dust-to-gas ratio in the DLA toward 
AO0235+164 is estimated to be 0.19 times the Galactic value (Junkkarinen 
et al. 2004). On the other hand, the observed feature is stronger 
than expected from $E(B-V) = 0.23 \pm 0.01$ for this absorber (Junkkarinen et al. 2004). 
In Galactic diffuse interstellar clouds, the peak optical depth in the 
9.7 $\mu$m silicate feature ($\tau_{9.7}$) is observed to correlate with the 
reddening along the line of sight, with $\tau_{9.7} = A_{V}/18.5$ (e.g., Whittet 1987). 
Extrapolating this relation, and assuming $R_{V} = 3.1$, one would 
expect $\tau_{9.7} \approx 0.039$ for the DLA in 
AO0235+164. Our observations, however, indicate $\tau_{9.7} \approx 
0.08$ for this DLA, $\sim 2$ times higher 
than expected from the relation for Galactic diffuse ISM. 
The dust in this absorber may thus be somewhat richer in silicates 
than typical Galactic dust. We note, however, that the silicate 
feature is also known to be stronger in the Galactic Center region, perhaps 
due to fewer carbon stars (and thus less carbonaceous dust) there (e.g., 
Roche \& Aitken 1985). If future observations of other DLAs also reveal 
material richer in 
silicates, it might indicate that those DLAs probe denser regions near 
the centers of the respective galaxies. 

The Galactic interstellar 9.7 $\mu$m feature is 
generally broad and relatively featureless, which is taken as an indication 
that interstellar silicates are largely amorphous. 
(Crystalline silicates would produce structure within the broad feature.) 
In principle, silicate grains may be composed of a mixture of 
pyroxene-like [(Mg$_{x}$Fe$_{1-x}$)SiO$_{3}$] and olivine-like 
[(Mg$_{x}$Fe$_{1-x}$)$_{2}$SiO$_{4}$] silicates, with the shape and 
central wavelength of the 9.7 $\mu$m absorption somewhat dependent 
on the exact composition (e.g., Kemper et al. 2004; Chiar \& Tielens 2006). 
Fig. 1(b) shows a closer view of the data, normalized by the power law 
continuum shown in Fig. 1(a), and binned by a factor of 3. 
The dotted and short-dashed curves are fits based on silicate emissivities  
derived from observations of the M supergiant $\mu$ Cep and of the 
Orion Trapezium 
region (e.g., Roche \& Aitken 1984; 
Hanner et al. 1995), which are taken to be representative of diffuse Galactic 
ISM and denser 
molecular material, respectively . The long-dashed and dot-dashed  
curves are fits based on the silicate absorption profile observed 
toward the Galactic Center Source 
GCS3, and on laboratory measurements for amorphous olivine (Spoon et al. 2006). 
The shape of the silicate profile observed toward AO 0235+164 is most similar 
to that of 
laboratory amorphous olivine, but the $\mu$-Cep and GCS3 templates also 
yield reasonable fits. The DLA silicate profile does not exhibit the 
redward extension seen for the Trapezium profile, suggesting that 
the DLA dust resembles dust in diffuse Galactic clouds more than 
that in molecular clouds. Using  $\chi^{2}$ minimization 
for 8.0-13.3 $\mu$m in the DLA rest frame, the peak optical 
depth values $\tau_{9.7}$ 
for the laboratory olivine, GCS3, $\mu$ Cep, and Trapezium templates are $0.081 \pm 0.018$, 
$0.088\pm 0.020$, $0.083 \pm 0.018$, and $0.071 \pm 0.016$, 
respectively for the binned data ($0.081 \pm 0.020$, $0.091 \pm 0.023$, 
$0.084 \pm 0.021$, and $0.069 \pm 0.017$, respectively, for the 
unbinned data). 
The error bars on $\tau_{9.7}$ correspond to optical depths that give 
reduced $\chi^{2}$ larger by 1.0 than the minimum values. The 
respective reduced $\chi^{2}$ values are 1.22, 1.32, 1.51, and 1.92 for the 
binned data (1.82, 2.08, 2.10, and 2.65 for the unbinned data). It is 
interesting to note that the best-fit astronomical template is GCS3, 
consistent with the enhanced $\tau_{9.7}/E(B-V)$ ratio seen in the DLA as 
toward the Galactic center.  
While the minimum reduced $\chi^{2}$ values are greater than 1.0, 
they are similar to those found in other 
studies of the silicate absorption toward both Galactic and extragalactic 
sources (e.g., Hanner et al. 1995; Bowey et al. 1998; Roche et al. 2006, 
2007). Indeed, we do not expect a perfect fit, since possible differences in 
dust grain size and chemical composition can alter the shape of the silicate 
feature, including the peak wavelength and the FWHM (Bowey et al. 1998 and 
references therein).  Higher S/N and higher resolution data would be needed to 
shed further light on the specific types of silicates present in DLAs.

 With a larger absorber sample, it would be possible to explore 
correlations between the strengths of the 
9.7 $\mu$m silicate feature and the 2175 {\AA} extinction bump  
(which is thought to be produced by a carbonaceous 
component of the dust).  For example, it would be 
interesting to understand whether 
the relative amounts of silicate and carbonaceous dust vary with redshift  
or with the gas-phase abundances of C or Si. High-S/N observations of other 
possible features (e.g., the 18.5 $\mu$m silicate 
feature or the 3.0 $\mu$m H$_{2}$O ice feature) would provide additional 
constraints on  
dust composition. (While those features are generally weaker than the 9.7 
$\mu$m feature in the Milky Way, the 3.0 $\mu$m feature can be stronger than the 
9.7 $\mu$m feature in highly reddened molecular sightlines.) 

Our exploratory study has demonstrated the potential of the Spitzer IRS to study dust 
in quasar absorbers. It would be very interesting to obtain similar 
spectra for other dusty quasar absorbers. The $E(B-V)$ values for dusty 
absorbers such as that reported here (0.23) are much larger than those for  
typical Mg II absorbers  
[$E(B-V)$ of 0.002; York et al. 2006b]. These relatively large reddening 
values are comparable 
to some of those for Ly-break galaxies (LBGs), which show $E(B-V)$ up to 0.4 and a median 
$E(B-V)$ of $\approx 0.15$ at $z \sim 2$ and $z \sim 3$ 
(Shapley et al. 2001, 2005; Papovich et al. 2001). Such dusty absorbers 
appear to be chemically more evolved (Wild et al. 2006) than typical DLAs, and may 
possibly provide a link in terms of SFRs, masses, metallicities, and dust 
content between the primarily metal-poor and dust-poor general DLA 
population with low SFRs and the actively star-forming, metal-rich, and dust-rich 
LBGs. Further Spitzer IRS observations of more dusty quasar absorbers thus will help 
to open a new window on this interesting class of distant galaxies. 

\acknowledgments
This work is based on observations made with the Spitzer Space Telescope, which is 
operated by the Jet Propulsion Laboratory, California Institute of Technology under a contract 
with NASA. Support for this work was provided by NASA through an award issued 
by JPL/Caltech. VPK acknowledges support from NSF grant AST-0607739 to 
University of South Carolina. 
DEW acknowledges support from NASA LTSA grant NAG5-11413 to the University
of Chicago. 
We are grateful to the Spitzer Science 
Center staff for helpful advice on data analysis and to an anonymous referee for 
helpful comments. 


%% in parentheses, after the keyword, but they will not be verified.

{\it Facilities:} \facility{SST (IRS)}.

\begin{figure}
\epsscale{0.77}
\plottwo{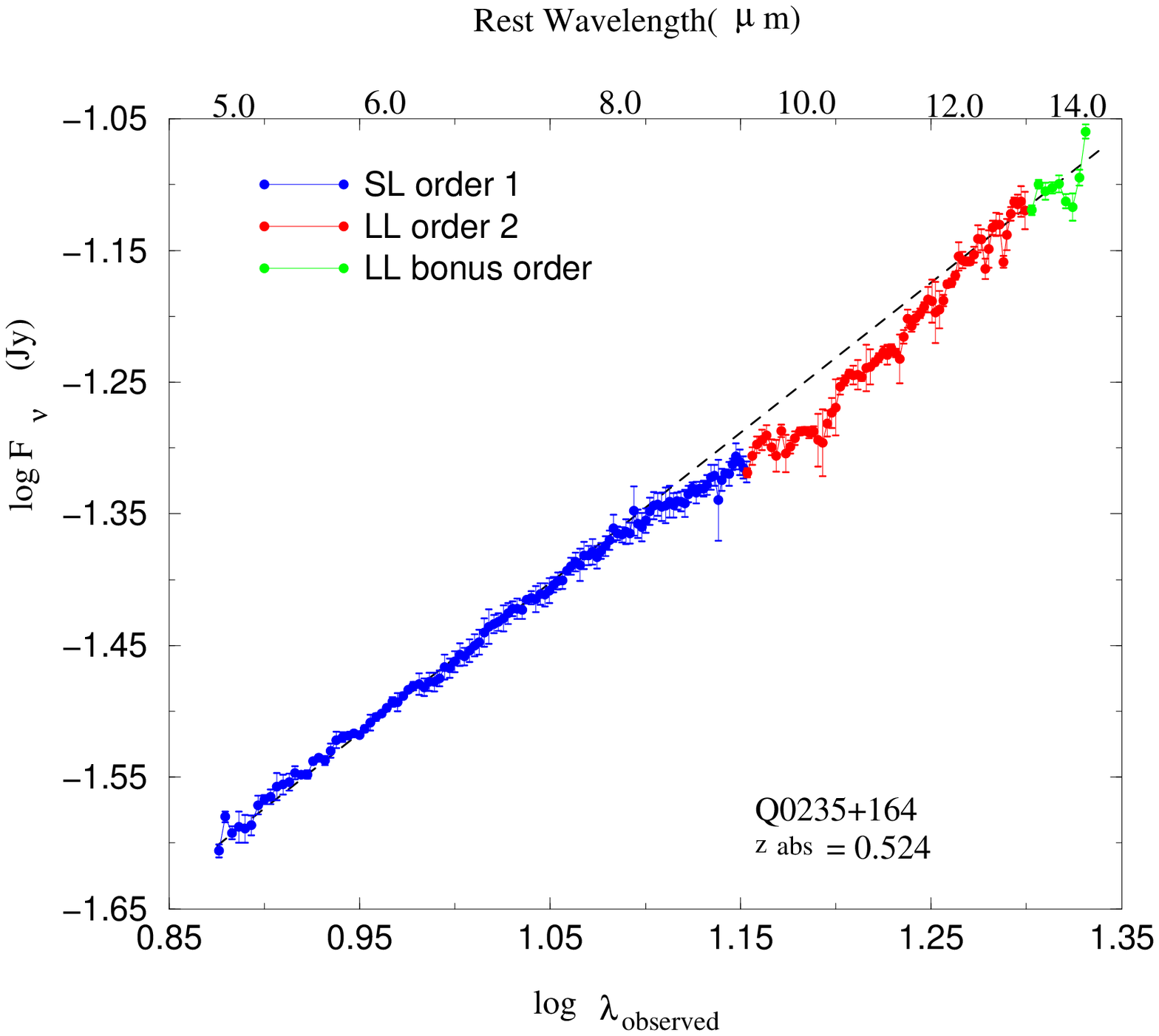}{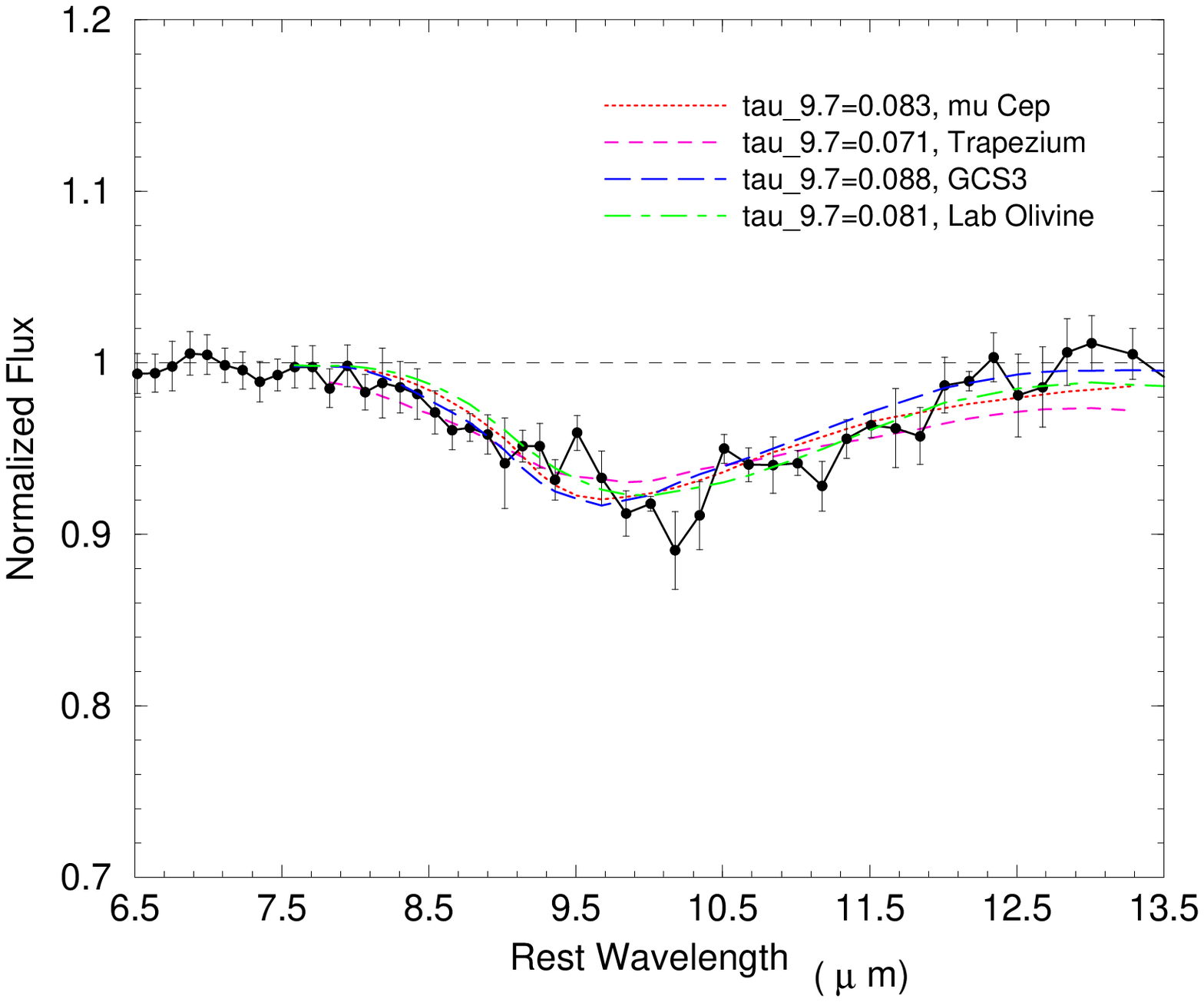}
\caption{(a) {\it Left:} Spitzer IRS spectrum of AO0235+164. The lower scale 
for 
the abscissa denotes the logarithm of the observed wavelength in $\mu$m;  
 rest frame wavelengths at the absorber redshift are shown at the top. 
The errorbars denote 1 $\sigma$ flux uncertainties. 
The dashed line shows a power law estimate of the 
continuum. (b) {\it Right:} A closer look at the suggested silicate feature. 
The abscissa denotes the rest frame wavelength at the DLA  
redshift. The data points show the spectrum, normalized by the 
power law continuum and binned by a factor of 3. The errorbars denote 
1 $\sigma$ uncertainties. 
The smooth curves show profiles for four templates of silicate 
optical depth, based on observations for 
three Galactic sightlines and laboratory measurements for amorphous olivine.} 
\end{figure}
%\clearpage
%
%% Here we use \plottwo to present two versions of the same figure,
%% one in black and white for print the other in RGB color
%% for online presentation. Note that the caption indicates
%% that a color version of the figure will be available online.
%%

%% This figure uses \includegraphics to scale and rotate the still frame

%% If you are not including electonic art with your submission, you may
%% mark up your captions using the \figcaption command. See the
%% No more than seven \figcaption commands are allowed per page,
%% so if you have more than seven captions, insert a \clearpage

\clearpage

\end{document}